 \documentclass[preprint2]{aastex}

\begin{document}


\title{Multiple Collimated Outflows in the Proto-planetary Nebula GL 618}


\author{Susan R. Trammell}
\affil{Physics Department, University of North Carolina at Charlotte,
    Charlotte, NC 28223}
\email{srtramme@email.uncc.edu}

\and

\author{Robert W. Goodrich}
\affil{The W. M. Keck Observatory, 65-1120 Mamalahoa Highway, Kamuela, HI 96743}
\email{rgoodrich@keck.hawaii.edu}


\begin{abstract}
We present narrow-band H$\alpha$, [\ion{S}{2}], and [\ion{O}{1}] Hubble Space Telescope images of the young planetary
nebula
GL 618. This object is a compact, bipolar nebula that is currently undergoing
the transition
from asymptotic giant branch star to planetary nebula. Our images confirm
the presence of
at least three highly collimated outflows emanating from the central regions
of GL 618. We
also detect H$\alpha$ emission close to the central dust lane and in an
extended scattered light
halo. The three outflows are occurring simultaneously in this object,
as opposed
to being the result of a precessing jet. We derive an inclination for
the brightest outflow in
the East lobe of 39\degr $\pm$ 4\degr. This differs from the previous
estimate of 45\degr. In
addition, our results indicate that the outflows seen in GL 618 are
probably not coplanar.
Line strengths derived from the narrow-band images indicate a shock
velocity in the range
of 40 $-$ 100 km s$^{-1}$. Based on the shock velocity we estimate
that the age of the
outflows is less than 500 years. The outflows seen in the optical
images of GL 618 are
related to features seen in near-IR, CO and CS maps of this object.
This relationship indicates that
the outflows are playing a major role in the morphological evolution
of this young planetary
nebula, interacting with and shaping the neutral envelope surrounding
GL 618. We discuss
the implications of these jets and their interaction with the neutral
envelope in the context of
current models of planetary nebula formation.
\end{abstract}




\section{Introduction}

\citet{kwo78} introduced the interacting winds model to account for the development of
spherical planetary nebulae (PN). In this model a massive, low-velocity wind removes a
large fraction of the envelope material during the asymptotic giant branch (AGB) phase of
evolution, and is followed by a higher-velocity, lower-mass wind. The high-velocity wind
snowplows into the previously ejected AGB wind material and sweeps up gas to form a PN.
The majority of PN exhibit aspherical structure, ranging from slightly elliptical to bipolar
\citep{gre71,zuc86,bal87,sch92,man96}. Other investigators have extended the interacting
winds model to explain the formation of these aspherical objects
\citep{kha85,bal87,sok89,ick92,fra94,dwa96}.

Understanding the origin of asymmetric structure in PN has been the focus of extensive
research in recent years. To achieve asymmetric PN morphologies, the interacting winds
model requires that mass loss during the early phases of the development of these objects is
aspherical. This aspherical structure is then amplified through the wind-shaping that occurs
as the PN evolves. Observationally, the time that aspherical structure becomes important is
becoming clear. Mass loss during the AGB is predominantly spherical
\citep{ner98,mei98},
while the morphologies of proto-planetary nebulae are typically aspherical
\citep{tra94,mei97,day98,kwo98,uet00}. These findings indicate that during the late-AGB
and/or post-AGB phase, the mass loss becomes aspherical \citep[for reviews]{kas00}. The
mechanism that triggers this change in the character of the mass loss is not well understood.
Suggestions for this trigger range from binary star evolution \citep{mor81,sok94} to the effects of
magnetic fields \citep{gar97}.

A recent HST narrow-band H$\alpha$ survey of the
morphologies of low excitation
PN \citep{sah98a} revealed a dazzling array of
complex structures in these
objects including jets, knots, bubbles, and loops.
The large variety of structures seen in these
images cannot be explained by
the interacting winds model alone.
Sahai \& Trauger suggested that phenomena such
as jets acting during the early post-AGB phase
may play a role in creating
the complex structures seen in these
objects. Recent CO observations of several
young PN with collimated outflows have demonstrated
that these outflows are significantly disrupting and
interacting with the neutral shells surrounding these objects \citep{for98,cox00,hug00,bac00}.
Both the HST imaging and the CO observations suggest
that jets could have a significant impact on
the way we think about the evolution of PN. The
interaction of the jet with previously ejected
material may influence the morphological evolution
of PN in a way that is not considered in current
models of PN formation. Further, the
phenomena driving the change from spherical to
aspherical mass loss in the early stages of
PN development are not clear. The formation of
jets may play a role in the change to
aspherical mass loss in some objects.

If jets are playing a major role in the shaping of PN, their influence will be most significant
at a time when the PN still has a large neutral envelope and before the fast wind develops.
We present HST observations of GL 618, a young PN undergoing the stage of evolution
immediately preceding the PN phase, during which recently lost material has begun to
disperse but the fast wind has not fully developed. GL 618 is surrounded by a shell of dust
and neutral gas that extends well beyond the optical nebula \citep[and references
therein]{spe00,mei98}. Our new observations have allowed us to confirm that jets are present
in GL 618 and to better understand the importance of these outflows in development of this
object.

Ground-based optical and near-IR imaging of GL 618 reveal two lobes of emission (each
about 3$\arcsec$ in extent) separated by a dark lane. The central regions of the nebula are
hidden from direct view at optical and near-IR wavelengths by the lane of obscuring
material. The spectrum of GL 618 is composed of a faint continuum and a variety of
low-excitation emission lines. \citet{sch81} and \citet{tra93} used spectropolarimetry to study
GL 618 and found that the continuum and part of the permitted line emission are reflected
from deep in the nebula. The low-excitation, forbidden line flux and remainder of the
permitted line emission are produced in the bipolar lobes. Trammell et al. demonstrated that
shock-excited emission dominates the spectra of the lobes of GL 618 with the emission
indicative of shock velocities in the range of 50$-$100 km s$^{-1}$. 
Long-slit optical
spectroscopy of GL 618 suggests that the shock emission is associated with out-flowing
gas \citep{car82}.
\citet{san02} present a recent detailed ground-based spectroscopic study of GL 618. Their analysis of the emission line profiles and ratios suggest shock velocities in the range
75$-$200 km s$^{-1}$. The largest velocities correspond to the tips of the
bullets that we see in our HST images.
Near-IR spectroscopy of GL 618 has revealed the presence of
thermally excited H$_2$ emission \citep{thr81,lat92}. This object also exhibits [\ion{Fe}{2}]
emission thought to be associated with the shock-heated gas \citep{kel92} and more
recent observations establish that at least part of this emission is associated with an outflow
\citep{kas01}. \citet{uet01} presented high spatial resolution near-IR images of GL 618
taken with the Subaru Telescope. The structure observed in the near-IR extends beyond the
optical nebula and these images reveal the presence of "horns" and bullet-like structures
associated with the bipolar lobes of GL 618. Ueta et al. suggest that the [\ion{Fe}{2}]
emission is associated with these features. These features are coincident with the tips of the
outflows that we see in our HST images. These authors also found high velocity CO
clumps at the positions of the IR bullets when they re-examined existing $^{12}$CO J=1-0
interferometry data. Our new HST observations demonstrate that the source of the shock-excited
emission in GL 618 is a set of highly collimated outflows originating in the central
regions of the object.

\section{Observations}

We have obtained WFPC2 images of GL 618 as part of a HST Cycle 6 program. GL 618
was centered in the Planetary Camera which has a 36\arcsec x 36\arcsec field of view and a
plate scale of 0.0455\arcsec pixel$^{-1}$. Images were obtained through three filters
selected to isolate strong shock-excited emission lines, and thus trace the outflow in this
source. The filters used were F631N (isolating [\ion{O}{1}]$\lambda$6300 line
emission), F656N (isolating H$\alpha$ line emission), and F673N (isolating [\ion{S}{2}]
$\lambda$$\lambda$6717,31 line emission). In addition, we obtained images through
F547M in an attempt to measure the adjacent continuum emission. The images of GL 618
were obtained on 23 October 1998 UT and exposure times ranged from 15 to 45 minutes.
The images were processed through the HST data reduction pipeline procedures and cosmic
rays were removed by combining several exposures of each object. The images were flux
calibrated using the standard conversion to flux provided during the pipeline reduction
procedures.

We have not removed the continuum from the narrow-band images that we present in this
work. The F547M continuum band is contaminated by many weak emission lines making it
impossible to reliably use images through this filter for continuum subtraction.
Alternatively, we could have used an average value for the continuum derived from our
ground-based work to remove continuum emission from our HST images. However, the
line to continuum ratio is highly variable across the nebula making this approach unreliable.
We point out that the continuum is approximately flat across the wavelength range covered
by our narrow-band filters (as indicated in the ground-based spectroscopy,
\citet{tra93})
and that this continuum is significantly weaker than the line emission in our ground-based
spectroscopy (average line to continuum ratios of 50 to 125 for H$\alpha$, [\ion{S}{2}],
and [\ion{O}{1}]). Further, the line emission seen in our HST images is concentrated in
knot-like structures. In these regions of intense line emission, the line to continuum ratio is
higher than the average values derived from the ground-based work. Consequently, the lack
of continuum subtraction does not significantly affect our conclusions regarding the overall
morphology of the shock-excited emission.

We have used our narrow-band images to form several line ratio maps for GL 618. The
problems with continuum subtraction do introduce an additional error (approximately 10\%
in the worst case) into these maps. However, the primary purpose of these maps is to
determine the outflow velocity. The outflow velocity is determined by comparing the
observed line ratios with ratios predicted by models of shock-excited emission. The
comparison with shock models results only in a range of possible shock velocities. The
relatively small errors introduced into the line ratios by the lack of continuum subtraction do
not significantly alter the value of the shock velocity that we derive.

\section{Results And Analysis}

Previous ground-based optical and near-IR work indicate that out-flowing gas is present in
GL 618 (see section 1 for discussion). We present HST narrow-band imaging of GL 618
that confirm the presence of outflows in the lobes of this object and reveal the morphology
of these flows. Figure 1 shows the 3-color image of GL 618 created by combining the
H$\alpha$, [\ion{S}{2}], and [\ion{O}{1}] narrow-band images of this object. Figure 2
shows the individual narrow-band images. At least three distinct outflows are visible in each
lobe of GL 618. The opening angles of these outflows are small, approximately 10 $-
$15\degr. The central star is not visible in the HST images, consistent with previous optical
and near-IR images \citep{lat92}. A dust lane obscures the central star at optical
wavelengths.

The morphologies seen in the [\ion{S}{2}] and [\ion{O}{1}] HST images are similar (see
Figures 1 and 2). These images trace the morphology of the shock-excited emission in GL
618. The brightest emission occurs near the tip of each of the outflows and there is no
forbidden line emission seen in the central regions of GL 618. Careful examination of the tip of
the outflow labeled {\it a} in the west lobe of GL 618 (see Figure 1)
reveals an excitation gradient across this region.
H$\alpha$ is brightest on the
side of the knot that is facing away from the central regions of GL 618. This type of
gradient is expected as material flowing away from the central source
impinges on the surrounding nebular material. The bright spots at the tips of the outflows
are not clumps of material being overrun by a wind or outflow \citep{tra00}. In this case, the brightest H$\alpha$ emission would be present on the side of the bullet-like structures facing the central object. Based on
their near-IR images and a reanalysis of high resolution CO data, \citet{uet01} also argue
that the shock-excitation is produced when fast-moving material impacts the surrounding
nebula.

The morphology observed in H$\alpha$ differs from the forbidden line morphology.
H$\alpha$ emission is associated with the outflows, but in addition, a significant amount of
H$\alpha$ emission is seen towards the central regions of GL 618 (see Figures 1 and 2).
Spectropolarimetric observations indicate that part of the H$\alpha$ emission is reflected
and part of this emission is produced by shocks in the lobes \citep{tra93}. We have
spatially separated these components in the HST images. At least part of the H$\alpha$
emission associated with the central regions of the object is probably reflected emission
from an \ion{H}{2} region buried deep in the nebula.
Recent ground-based spectroscopy confirms that part of the H$\alpha$ in
the inner regions is scattered \citep{san02}.
A high density \ion{H}{2} region
has been observed at the center of GL 618 at radio wavelengths \citep{kwo81} and in the
reflected optical spectrum \citep{tra93}. The H$\alpha$ emission coincident with the
outflows is the shock-excited component of the permitted line emission.

In addition to the H$\alpha$ emission associated with the inner regions of the object and
the outflows, a faint halo of emission is present in the H$\alpha$ image (best seen in Figure
2 panel {\it a}). This faint halo is most likely H$\alpha$ emission scattered by dust present in
the outer regions of GL 618. GL 618 is known to be surrounded by an extensive molecular
and dust envelope (see section 1 for discussion). The dusty material is probably part of the
shell of material ejected by the central star of the object during its late AGB evolution. The
axis of symmetry appears to be different for the scattered light envelope and the outflows
seen in GL 618.

Three outflow features are seen in the images of GL 618. We believe that these features are
the result of three separate outflows occurring simultaneously in GL 618. Alternatively,
these features could be the result of one precessing jet. However, if this were the case we
would expect to see S or spiral shaped morphology in the emission from the outflow
\citep{cli95,liv96}. We do not observe this type of morphology in the lobes of GL 618. We
have further investigated the nature of the outflows by considering the recombination times
of the ionized gas in the outflows. The presence of photoionized gas appears to be limited to
the immediate vicinity of the central star in GL 618. This suggests that there is no
significant source of ionization in the lobes of GL 618 other than that caused by thermal
excitation in the outflows. If an outflow precesses away from a region, that region will
continue to emit until the gas cools and recombines. Emission from S$^+$ is seen in all
three outflow features. We can estimate the amount of time that the gas would continue to
produce [\ion{S}{2}] line emission after the outflow precesses away from a region by
calculating the recombination time for S$^+$. The recombination time, $\tau_{recomb}$,
is given by
\begin{displaymath}
\tau_{recomb} = \frac{1}{N_e \alpha}
\end{displaymath}
where {\it N$_e$} is the electron density and $\alpha$ is the recombination coefficient.
The electron density is approximately 10$^4$ cm$^{-3}$ based on ground-based
spectroscopy \citep{tra93} and the recombination coefficient for S$^+$ is 1.8 x 10$^{-
12}$ cm$^3$ s$^{-1}$ \citep[for T=10,000 K]{arn85}. The S$^+$ would recombine in
approximately 1 year after the outflow moved away from a region. Based on this result, we
conclude that the three features seen in our HST images are three separate ongoing
outflows.
It has been suggested that simultaneous, multiple outflows have occurred or are occurring in several other young PN including the Starfish Twins \citep{sah00}, GL 2688 \citep{cox00}, Roberts 22 \citep{sah99}, and the Frosty Leo Nebula \citep{sah00a}.

A ring-like morphology is seen in the outflows in all of the HST images (see Figure 1).
This is most evident in the bright outflow labeled {\it b} in Figure 1. This morphology has
not been seen in any ground-based images. The presence of these structures offers an
opportunity to estimate the inclination of GL 618. \citet{car82} estimated the inclination of
GL 618 be to 45\degr based on their optical spectroscopy. If GL 618 were viewed pole-on
($i =$ 90\degr) we would expect the rings to be perfect circles. Instead, we see elliptical
structures. We have estimated the inclination of three of the outflows based on the shapes of
the rings as seen in the H$\alpha$ and [\ion{S}{2}] images. For the brightest outflow in
the East lobe (labeled {\it b} in Figure 1), we estimate an inclination of 39\degr $\pm$
4\degr. For the outflow labeled {\it c} in the East lobe, we estimate an inclination of 59\degr
$\pm$ 10\degr. The feature labeled {\it d} in Figure 1 implies an inclination of 36\degr
$\pm$ 4\degr. These results suggest that the outflows in GL 618 are not co-planar. These
results also indicate that the previous estimate of the inclination of GL 618 was an
overestimate.
\citet{san02} also find a lower inclination for GL 618. Based on their spectroscopic observations of H$\alpha$ in GL 618, these authors calculate
an inclination angle of 24\degr $\pm$ 6\degr.

The ring-like features occur at semi-regular intervals along the outflows and could indicate
that the source of the jet is episodic. \citet{rag98} modeled this type of
structure for episodic outflows associated with young stars.
Alternatively, the rings could be the result of
instabilities in the outflow. If the rings are the result of an episodic jet, we can estimate the
amount of time between ejection events by looking at the spacing of the rings. The ring-like
structures seen in outflow labeled {\it b} in Figure 1 are separated by approximately
0.5\arcsec. Assuming an outflow velocity of 120 km s$^{-1}$ (see discussion later in this
section) for the jet and a distance to GL 618 of 1.5 kpc, this implies an ejection event
approximately every 35 years. Note that distance to GL 618 is uncertain and estimates range
from 0.9 to 2 kpc \citep{sch81,goo91}.

A dark arc is evident in the H$\alpha$ image of the West lobe (see Figure 2a). This arc is
located approximately 1.5\arcsec from the center of the dark lane. The dark arc is a region where there is a deficit of
H$\alpha$ emission. We believe that we are seeing dust in
silhouette at the position of the dark arc.
This dust could be located in the
outer edge of the lane of dust obscuring the central star of GL 618. On the other hand, it could be a feature in
the extended halo of GL 618. This halo is the remnant of AGB mass loss in this object. If the arc is a feature of the dust halo of GL 618, it might be
similar to the arcs seen in HST images of GL 2688, another young planetary nebula. The
arc structure in the extended scattered light halo of GL 2688 is explained as episodic mass
loss by the star during the late AGB phase of evolution \citep{sah98b}. A deep image of the
scattered light halo of GL 618 could reveal if this type of episodic mass loss occurred in
this object.

We have measured the average value of line ratios at the tips of the outflows labeled {\it a, b,}
and {\it e} in Figure 1. These are the regions where out-flowing material is impacting the
surrounding nebula.
In these regions
[\ion{O}{1}] / H$\alpha$ = 0.37 $-$ 0.40 and [\ion{S}{2}]/H$\alpha$ = 0.11 $-$ 0.16.
The values of the line ratios at the tips of the outflows provide an
estimate of the shock velocity, and thus the outflow velocity in these regions.
The bullet-like structures seen at the tips of the outflows have the geometry of bow shocks.
Our estimates
are an average of the line ratios across the entire bow shock region. Only the line ratios at the head of the bow shock, where the velocity component perpendicular to the shock is greatest, will reveal the maximum shock velocity. The rest of the bow shock consists of oblique shocks that result in line ratios indicative of lower shock speeds. Therefore, the shock velocity derived by examining line ratios from the entire region of the bow shock will provide an average value of the shock velocity. The outflow velocity at the tip of the outflow will be larger.

We have compared
the observed line ratios with the models of thermally excited emission produced in a bow
shock presented by \citet{har87}. The line ratios are most consistent with their model of a
bow shock with velocity 100 km s$^{-1}$.
If we compare our results to the planar shock
models presented by Hartigan et al., we find that our observed line ratios are consistent with
a shock velocity in the range of 40$-$100 km s$^{-1}$. Note that these models do not include the effects of magnetic fields. The field strength at the tips of the outflows is likely to be small so that this assumption is valid. However, if the field is strong ($\geq$ 300 $\mu$G) in these regions, we could be underestimating the shock velocity by 20$-$40 km s$^{-1}$.

The shock velocity is a measure of the relative speeds of the colliding gas. In the case of GL
618 an outflow is colliding with a surrounding shell of material that was ejected during the
AGB/post-AGB phase. \citet{mei98} have estimated an overall outflow velocity for the
extended CO shell around GL 618 of 20 km s$^{-1}$. Assuming that the outflows are
colliding with gas moving at approximately 20 km s$^{-1}$, the measured shock velocity
(100 km s$^{-1}$) indicates a velocity for the collimated outflow to be about 120 km
s$^{-1}$ (V$_{outflow}$ = V$_{shock}$ + V$_{shell}$). This is consistent with
estimates of the outflow velocity based on previous spectropolarimetry \citep{tra93} and
with optical and near-IR kinematic measurements (see section 1).
This value is also consistent with the shock velocities recently reported by \citep{san02}.

Based on an outflow velocity of 120 km s$^{-1}$ and assuming a distance to GL 618 of
1.5 kpc, we calculate the kinematic age of the outflows. The brightest outflows (labeled {\it
a} and {\it b} in Figure 1) extend approximately 6\arcsec from the center of GL 618 in the
HST images. We assume an inclination of 39\degr (see previous discussion). These
parameters yield a kinematic age of $\lesssim$ 500 years for the outflows seen in the HST
images. The outflows seen in GL 618 are a recent phenomenon and probably became active
shortly after the star left the AGB.

\citet{haj95} detected emission from CS in GL 618. This emission has an X-like
morphology and is confined to the inner 4\arcsec of the object (this corresponds to the dark
lane that we see in our optical images). The outflows that we see in the optical seem to be a
continuation of the CS structure, as the optical outflows continue along a line extended from
the CS structure. \citet{mei98} examined the CO emission in GL 618 at high spatial
resolution. The majority of the CO emitting gas in GL 618 is participating in a spherical
expansion at 20 km s$^{-1}$. However, these high spatial resolution observations reveal
that the halo of CO emission is not entirely spherically symmetric. The CO exhibits an X-
like structure in the halo. This structure appears as 3 bright spots of CO emission located
approximately 20\arcsec from the center of GL 618. These areas of low-velocity (20 km
s$^{-1}$) CO emission are along the same axis as the CS and optical emission and are
exterior to the optical outflows. The fact that the low-velocity CO emission related to the
outflow is exterior to the optical emission is important. This indicates that the outflow is
interacting with the neutral shell surrounding GL 618. Jets are playing a major role in the
morphological development of this PN.

\section{Discussion}

Our observations demonstrate that emission from highly collimated outflows dominates the
optical emission and morphology of GL 618. A comparison of the optical images with
previous molecular line studies indicates that the jets are interacting with the neutral
envelope surrounding GL 618. This interaction is shaping the neutral envelope of GL 618
and will determine the morphological evolution of this object. GL 618 is undergoing the
early stages of PN development. Only a small region of photoionized gas is present in GL
618 and this ionized gas is found only in the immediate vicinity of the central star
\citep{kwo81}. The fast wind described in the interacting winds model \citep{kwo78} has
not fully developed in GL 618. The interacting winds model suggests that it is the
interaction of this fast wind with the surrounding neutral envelope that dictates the
morphological evolution of a PN. In GL 618 the fast wind will be interacting with a neutral
envelope that has already been significantly altered by the action of the jets.

The interaction of the jets with the neutral envelope seen in GL 618 is not unique. Recent
observations of several other young PN indicate that jets may have a much more significant
role in the overall development of these objects. CO observations of PN known to contain
collimated outflows have shown that these outflows are having a major impact on the
surrounding neutral shells \citep[e.g.]{for98,cox00,hug00,bac00}. For example, in BD+30\degr3639 the CO observations reveal a pair of high velocity molecular knots just outside the ionized nebula. This is very similar to relationship between the molecular material and optical jets seen in GL 618.
In all of these objects the
jet's interaction with the neutral shell, not interacting winds, could be the dominant
mechanism in determining the overall morphology of the PN.

The complex, multi-polar geometry observed in GL 618 is not unique. \citet{sah00} presented HST images of two young PN, He 2-47 and M1-37 (dubbed the Starfish Twins). These objects exhibit multi-polar morphology that is similar to that seen in GL 618. Sahai suggests that jets established the complex geometry of these nebulae. He estimates that the jets were active within the past few hundred years and suggests that multiple outflows occurred simultaneously in these objects. While the overall morphology of these objects is similar to GL 618, there are some differences. Knots or bullets of emission are prominent at the tips of the outflows in GL 618. These types of structures are not evident in the images of He 2-47 and M1-37. These bright spots mark the point where active jets collide with ambient material in GL 618. The jets are no longer active in the Starfish Twins so that the regions that correspond to tips have cooled significantly and do not appear as bright spots. 

CO observations and HST imaging strongly suggest that jets play a important or dominant role in the shaping of some PN. These jets appear to be short-lived and they seem to operate during the post-AGB and/or PPN phases of evolution. Early jet action sets the stage for the complex morphologies we observed in some PN. High spatial resolution spectroscopy of these objects is needed to better understand the dynamics of the outflows and their impact of the surrounding material.

The overall morphology of the neutral envelope of GL 618, as traced by CO, is spherical
\citep{mei98}. The more recent mass outflows in this object are clearly aspherical. As is the
case in other PN, the mass loss in GL 618 appears to have switched from spherical to
aspherical soon after the star left the AGB. The kinematic age of the outflows suggest that
the jets turned on sometime shortly after the star left the AGB. The trigger for the switch
from spherical to aspherical mass loss in GL 618 appears to be the formation of collimated
outflows in this object.

The debate concerning the origin of collimated outflows and also the formation of
aspherical PN in general, centers on whether binary or single stars are responsible for
producing aspherical mass loss. Both models of binary star interaction
\citep{mor81,sok94} and magnetic confinement \citep{gar97,mat00}, while providing
a scheme for producing the overall aspherical structure in PN, may also provide
mechanisms to produce the highly collimated outflows. The complex, multi-polar outflow
geometry seen in GL 618 may be difficult for either of these types of models to explain.



\acknowledgments
The authors thank the referee for useful comments and suggestions.
This work was supported by NASA through grant number GO-06761 from the Space
Telescope Science Institute, which is operated by AURA, Inc., under NASA
contract NAS 5-26555.




\clearpage


\begin{figure}
\epsscale{0.65}
\plotone{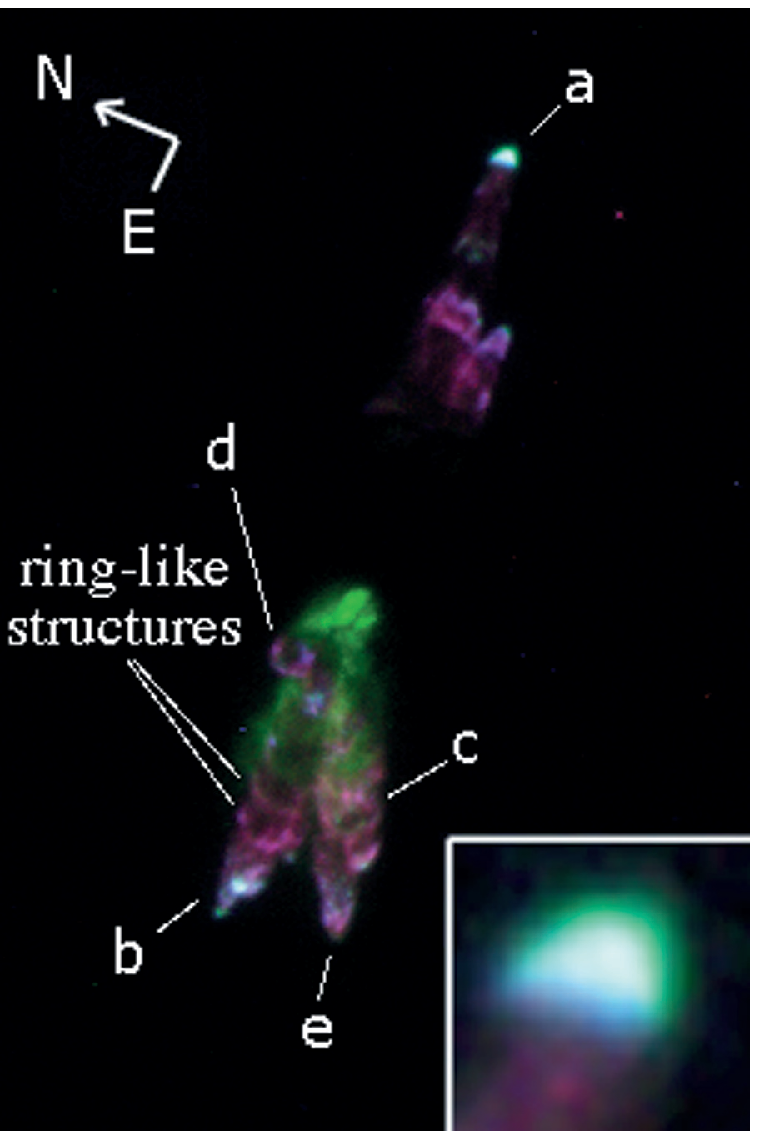}
\caption{Combination of three narrow-band WFPC2 images of GL 618. In this figure
green represents the H$\alpha$ emission, blue the [\ion{O}{1}] emission
and red the [\ion{S}{2}] emission. The forbidden line emission is
concentrated in three distinct collimated outflows that originate
in the central regions of GL 618. The H$\alpha$
emission is found in the outflows and components of this emission are
also present closer to the central dark lane of the object and in
in an extended halo. The size of this image is approximately 12\arcsec x
17\arcsec. The insert displays a detailed
view of the outflow labeled {\it a}. \label{fig1}}
\end{figure}

\clearpage 

\begin{figure}
\epsscale{1.0}
\plotone{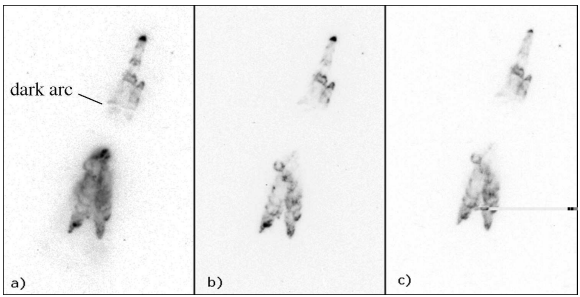}
\caption{(a) The F656N filter image which traces the H$\alpha$ emission in GL 618.
Notice the extended halo and the dark arc. (b) The F631N filter image
which traces the [\ion{O}{1}] emission (c) The F673N
filter image which traces the [\ion{S}{2}] emission. The size of each image
is approximately 12\arcsec x 17\arcsec. \label{fig2}}
\end{figure}

\end{document}